\documentclass[runningheads,a4paper]{style/llncs}
\usepackage{graphicx} 
\usepackage{amsmath}
\usepackage{geometry}
\usepackage{todonotes}
\usepackage{makecell}
\usepackage{authblk}
\usepackage[hidelinks]{hyperref}
\usepackage{placeins}
\usepackage[ruled]{algorithm2e}
\usepackage{changepage}
\usepackage[numbers,sort&compress,merge,round]{natbib}
\RequirePackage{jabbrv}
\usepackage{multicol}

\usepackage{graphicx} 
\usepackage{amsmath}
\usepackage{makecell}

\begin{document}

\title{How opinions get more extreme in an age of information abundance}

\author{
    Guillaume Deffuant\inst{1,4} \and 
    Marijn A. Keijzer\inst{2} \and 
    Sven Banisch\inst{3}}
\institute{
    Universit\'e Clermont-Auvergne, Inrae, UR LISC / LAPSCO, France \and
    Institute for Advanced Study in Toulouse / Toulouse School of Economics, France \and
    Karlsruhe Institute of Technology, Germany \and
    Corresponding author. E-mail: \href{mailto:guillaume.deffuant@inrae.fr}{guillaume.deffuant@inrae.fr}
}
\authorrunning{Deffuant, Keijzer \& Banisch}



\maketitle

\centerline{\normalsize Preprint; compiled on \today}

\begin{abstract}\normalsize
We live in an age of information abundance but know little about how this influences our opinions or attitudes. A common expectation is that people consulting numerous pieces of information, well balancing the different sides of an issue, will adopt a moderate attitude about the issue. We claim that this expectation is deceitful and suggest that people tend to get extreme and dogmatic about an issue when they consult abundant unbiased information. The cause for this extremization is a hardening confirmation bias---when their attitude gets more extreme, people get more likely to ignore information that differs from their views. Our claim is based on simulations of two fundamentally different computational models: a Bounded Confidence model and an empirically calibrated Persuasive Argument model. For both models, the attitude tends to be extreme when the computational agent consults abundant unbiased information. We analyze the extremization pathways displayed in the models and discuss how our results may affect views on polarization, and on the role of online media. 
\end{abstract}

\begin{adjustwidth}{3em}{3em}


\end{adjustwidth}

\vspace{3em}
\section*{Introduction}
The early 21st century marked a transformative era in the production, dissemination, and accessibility of information. Catalyzed by advancements in information technology---most notably, the proliferation of accessible content on the internet---large quantities of information became publicly and globally available. This ``information age’’ promised to democratize knowledge, and create well-informed citizens~\cite{benkler2002FreedomCommonsPolitical}. As individuals navigate these wider information spaces, its effects on political knowledge and beliefs may have critical consequences on the functioning of democracies~\cite{ guess2021ConsequencesOnlinePartisan,sunstein2002republic}.

When consulting a large number of information items presenting different sides of an issue, rational agents are expected to adopt a balanced or moderate attitude about the issue. According to this view, as the advent of the Internet made a huge amount of information about almost any subject readily available, we should live in a time of more appeased and moderate attitudes. However, there is evidence that, on the contrary, opinions about various societal issues of the elites and politically engaged people in the USA became increasingly extreme and entrenched over the last decade~\cite{McCarty2019}. 


Many people are convinced that online media have aggravated opinion polarization~\cite{lelkes2017hostile,fletcher2020HowPolarizedAre,melnikov2021MobileInternetPolitical} by creating personalized filter bubbles and ideological echo chambers~\cite{keijzer2022complex,Banisch2022modelling}, preventing users from integrating different sides of issues. Though evidence of this process is, at best, weak~\cite{budak_misunderstanding_2024,guess2023HowSocialMedia,guess2021ConsequencesOnlinePartisan,bruns2019filter,flaxman2016filter,zuiderveen2016should}, a common view is that online media platforms should expose their users to more diverse information in order to favor more balanced and well-informed attitudes, hence decreasing societal polarization. 
 
Other researches explain polarization as the effect of a negative influence (also called boomerang or reactance effect~\cite{Hovland1953}), according to which, people encountering views that are very opposed to theirs tend to increase their opposition to the received message instead of compromising~\cite{Keijzer2024}. Several computational models that assume negative influence predict more polarization when agents interact with more diverse opinions~\cite{Martins2010,Flache2011a,Axelrod2021}. However, empirical evidence about negative influence is mixed. It appears in the votes  of US senators~\cite{Liu2015} and in an experiment on Twitter~\cite{Bail2018}, while other experiments detect only weak negative influence (when participants already hold extreme views)~\cite{Keijzer2024} or fail to detect any~\cite{Hovland1957,Takacs2016}. 

In this paper, we argue that confirmation bias alone is sufficient to generate extremization of attitudes when consulting abundant and diverse information.
Confirmation bias expresses the general tendency to give attention, weight or credibility to information supporting our current attitude and to ignore or dismiss information contradicting our current attitude~\cite{nickerson1998confirmation,kappes2020confirmation,Shamon2019changing}. What is more, confirmation bias varies with the extremity of the current attitude. People with extreme attitudes tend to hold their attitude very firmly, while people with more moderate attitudes tend to be more open to change. In other words, confirmation bias tends to harden with attitude extremity~\cite{Hovland1980}. This feature of confirmation bias is well-supported experimentally~\cite{costello2023absolute}, but only few opinion dynamics models take it into account~\cite{Gargiulo2008,Banisch2023biased}. While numerous theoretical and empirical studies address the effect of confirmation bias on selection of information in social media~\cite{knobloch2020confirmation,pearson2019confirmation,pennycook2019lazy,goette2020information}, 
as far as we know, none of them investigates the mere effect of information abundance when confirmation bias hardens with attitude extremity. 

Our research draws on two different theoretical approaches to model how hardening confirmation bias and information abundance affect attitude extremization: the bounded confidence (BC)~\cite{Deffuant2000,hegselmann2002opinion} and persuasive arguments (PA)~\cite{myers1982polarizing,Banisch2021argument,Maes2013a} models.
Both models independently produce extreme attitudes when agents consult abundant unbiased information. 
The sheer breadth and variety of consulted content about a subject combined with hardening confirmation bias fuels a process of attitude extremization.



This process of attitude extremization is thus fundamental, as it is deeply rooted in our cognitive functioning.
In the current age of information abundance,
it could open new and unexpected pathways to extremization. 
Our work is the first to identify this process.

\section*{Model 1: agent with Bounded Confidence}

\begin{figure*}
\begin{adjustwidth}{-10em}{-10em}
    \centering
    \includegraphics[]{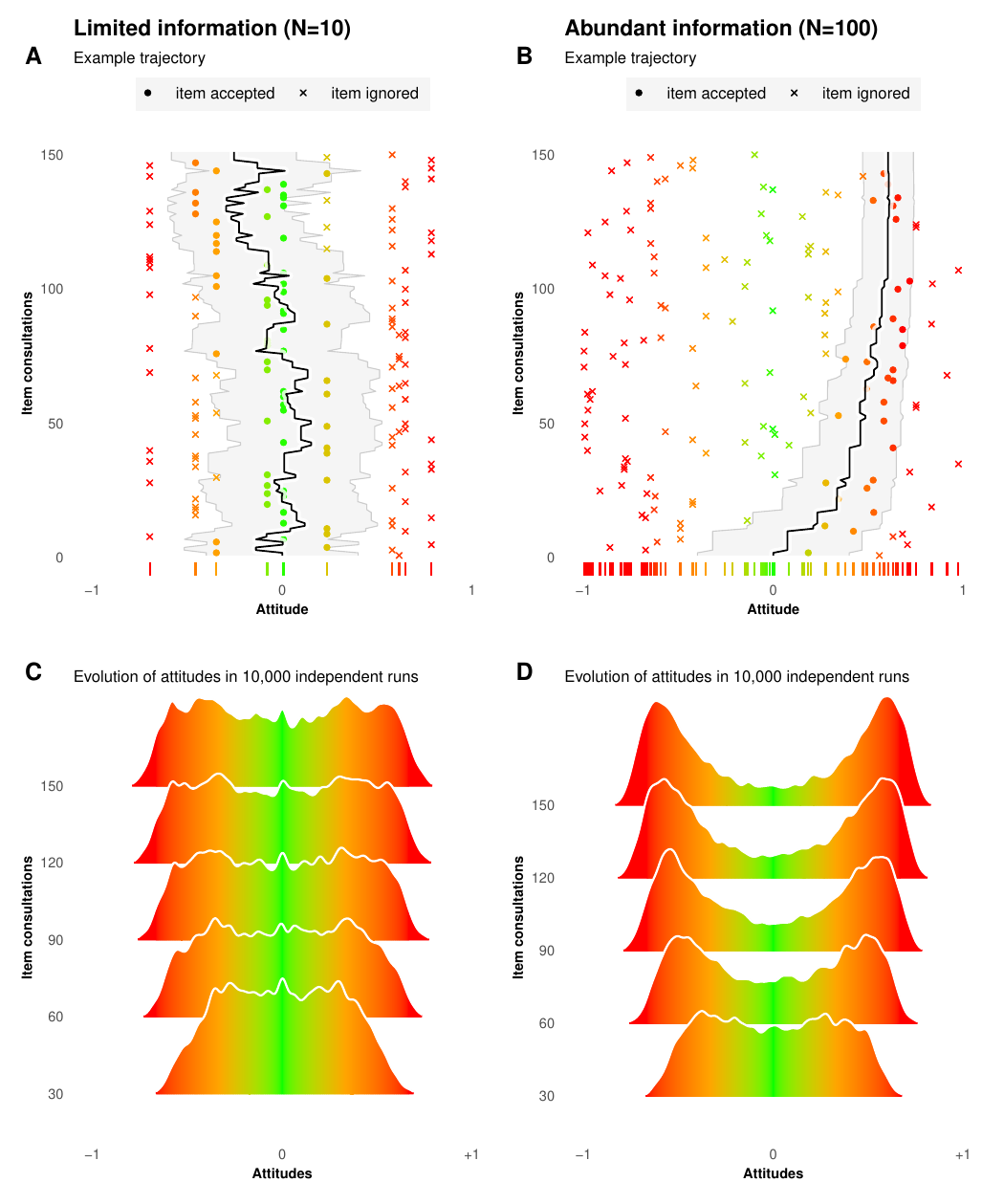}
\end{adjustwidth}
    \caption{Bounded Confidence model with a maximum confidence bound $\epsilon_M = 0.4$ and hardening bias parameter $\beta = 3$. Panels \textbf{A} and \textbf{B} show a trajectory of the attitude over 150 item consultations (black curve) and the corresponding confidence interval (in grey). The available items appear at the bottom of each panel. On panel A, with $10$ items, the attitude remain moderate because of gaps in the distribution of items. On panel B, with $100$ items, the gaps in the distribution of items are smaller and the attitude can more easily become extreme and then tends to stabilize because the confidence interval shrinks. Panels \textbf{C} and \textbf{D} show the distribution of attitudes derived from 10,000 runs for 150 item consultations, starting each time with a new set of items uniformly drawn in $[-1,1]$. Panel C shows the results with 10 items, and panel D shows the results with 100 items.}
    \label{fig:bounded-confidence}
\end{figure*}

Bounded Confidence (BC) models~\cite{Deffuant2000,hegselmann2002opinion} assume agents that communicate their attitude to each other. When receiving the attitude of another agent, agents change their attitude a little in the direction of the received attitude, only when the received attitude is close enough to their own. Agents ignore attitudes that are outside of their confidence interval and therefore `too far' from theirs. These models hence include confirmation bias at their core. The original version of the model assumes that all agents hold the same confidence bound~\cite{Deffuant2000}. Later variations of the BC model include radical agents, as characterized by extreme and persistent attitudes in some cases modeled by agents holding a narrow confidence bound~\cite{Deffuant2002,Deffuant2006,hegselmann2015OpinionDynamicsInfluence,keijzer2021StrengthWeakBots}. Here, we extend the BC model with the assumption that confidence bound decreases, hence confirmation bias hardens, when attitude extremity increases. 

\subsection*{Single agent model consulting randomly chosen items}

We consider a single BC agent with an attitude $a \in [-1,1]$ and $N$ available items of information about a topic (peer messages, news articles, videos, etc.). We assume that each item conveys an attitude about the topic, that we call the attitude of the item. 
In the model, we assume that the attitudes of the items are drawn from a uniform distribution on $[-1, 1]$. This assumption is conservative, as there is evidence that online, extreme attitudes are more frequent than moderate ones~\cite{Bronner2015}. The dynamics of the model repeats iterations in which the agent consults (i.e., reads or watches) an item $i \in \{1,...,N\}$ chosen at random and updates its attitude. If the item's attitude $c_i$ falls outside of the agent's confidence interval ($|a-c_i| > \epsilon(a)$, where $\epsilon(a)$ is the agent's confidence bound), the agent ignores the item and does not change its attitude. If, however $a$ and $c_i$ are close enough ($|a-c_i| \leq \epsilon(a)$), the agent accepts the item and adjusts its attitude in the direction of $c_i$. 

Crucially, since the attitudes of items are drawn from a uniform distribution on $[-1,1]$, the average attitude of the items is neutral (0). An unbiased agent consulting these items would therefore hold a neutral attitude on average.

\subsection*{Hardening confirmation bias}

Hardening confirmation bias is implemented as a shrinking confidence bound $\epsilon(a)$ when attitude extremity $|a|$ increases. It is maximal $\epsilon(0) = \epsilon_M$ at neutral attitude $a = 0$ and it is minimum at $\epsilon(\pm 1) = \epsilon_m$ when the attitude is the most extreme ($|a| = 1$). We assume that function $\epsilon(a)$ has a bell shape as suggested in empirical studies~\cite{costello2023absolute,Banisch2022validating}, determined by parameter $\beta$ (see Material and Methods for details) formally defined as
\begin{equation}
\epsilon(a) = \epsilon_M \exp(- \beta a^2)    
\end{equation}
where $\beta$ rules the hardening of confirmation bias. 
In the following computational experiments, the maximal confidence bound is $\epsilon_M=0.4$ and bias hardening parameter is $\beta=3$. In the SI, we show that the results are robust when changing these parameters.

\subsection*{Limited vs. abundant information}

We compare two experimental settings. In one setting, the number of available items about a topic is limited ($N=10$ in our experiments) representing the typical information offer of traditional media. In the other, this number of items is considerably larger ($N=100$ in our experiments), representing the typical offer of online media. Our aim is to determine if this change of information availability has an impact on the extremity of attitudes.

\subsection*{Results}

Fig.~\ref{fig:bounded-confidence} reports the results of a series of computational experiments for the BC model in settings of limited and abundant information. Panels A and B show examples of trajectories of the BC agents' attitude over $150$ item consultations, starting from a neutral attitude ($a_0 = 0$). The trajectory shown on panel A (obtained with limited information) fluctuates while remaining moderate, while the trajectory shown on panel B (obtained with abundant information) reaches a rather extreme attitude after about 100 item consultations and remains almost stable afterwards. 

Panels C and D of Fig.~\ref{fig:bounded-confidence} show the evolution of the distribution of attitudes computed from $10,000$ repetitions of 150 item consultations, drawing a new random set of items at the start or each run. When information is limited (panel~C), the distribution of attitudes tends to spread over time, but the moderate attitudes remain frequent. When information is abundant (panels~D), the frequency of moderate attitudes decreases when the number of consultations increases and extreme attitudes (extremity higher than 0.5) become considerably more frequent than when information is limited. 

A closer look to panels A and B of Fig.\ref{fig:bounded-confidence} exposes the mechanism responsible for the differences between limited and abundant information. 
On panel~A, the attitude remains moderate because the extreme items remain outside the confidence interval (shown in gray) of the agent. This situation is actually rather frequent because when drawing a small number of items uniformly, gaps of the order of the confidence bound between items are likely and the extremities of the confidence bound fluctuate in these gaps, preventing the confidence interval to include new extreme items.
By contrast, when the number of items is large (Panel~B), when the attitude becomes more extreme, the confidence interval is likely to include new extreme items, making possible further extremization. Then, once the agent extremizes slightly, its confidence interval shrinks and tends to exclude previously included moderate items. Hence, the agent is less likely to get back to previous moderate attitudes. In this way, 
extremization takes place simply because more extreme attitudes are more stable. Once the agent reached an extreme attitude, it tends to keep it. 
This process is caused by hardening of the confirmation bias and vanishes when confirmation bias is constant (see SI).

Moreover, it is important to notice that the process of extremization obviously requires extreme content to be available, otherwise the drift to extreme cannot take place. As expected, when extreme items are less likely, simulations with abundant information show mostly moderate attitudes (see details in SI). Therefore, the BC model suggests that decreasing the frequency of extreme content should mitigate extremization of attitudes due to abundant information.


\section*{Model 2: agent facing Persuasive Arguments}

\begin{figure*}
\begin{adjustwidth}{-10em}{-10em}
    \centering
    \includegraphics[]{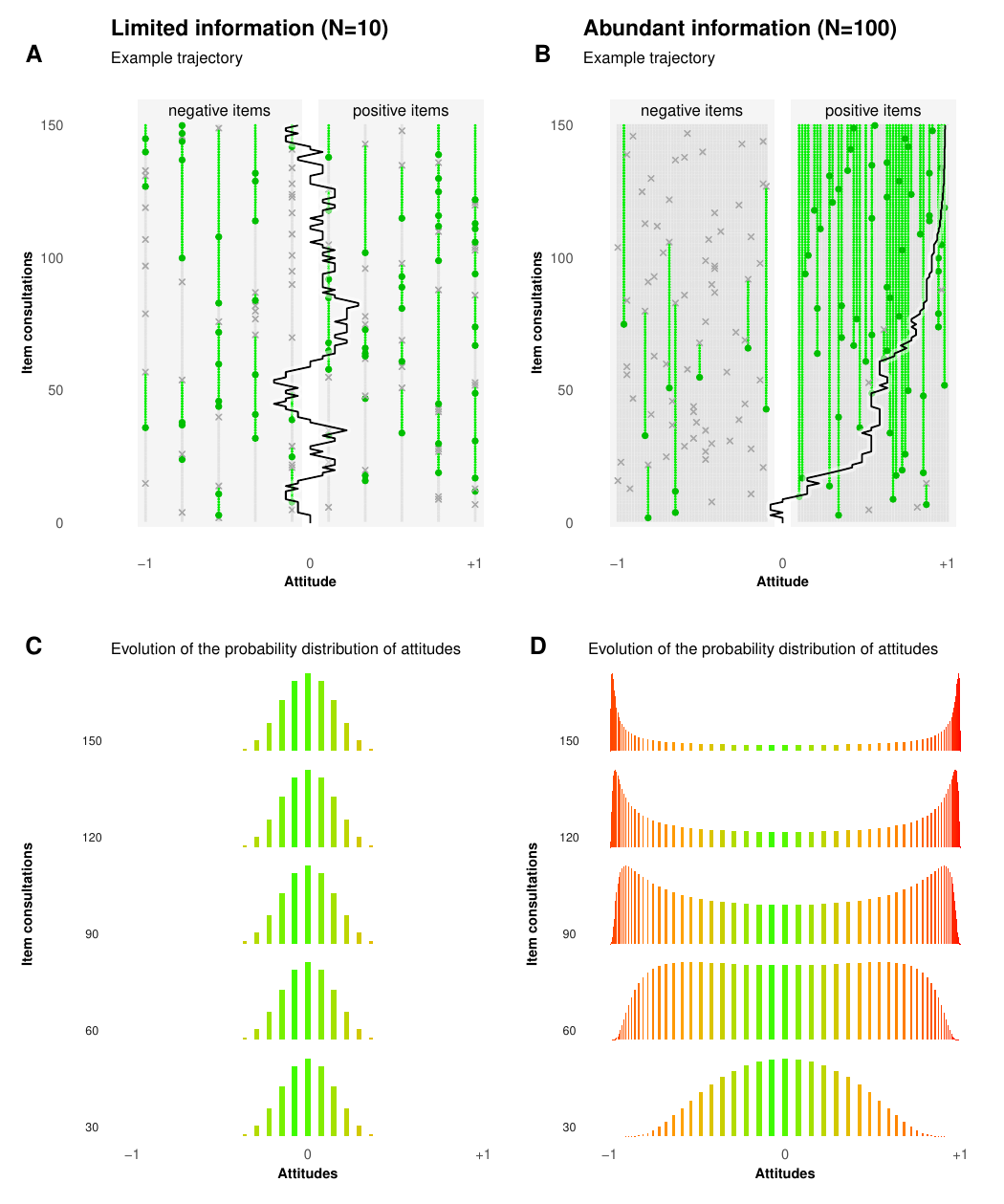}
\end{adjustwidth}
    \caption{PA model with hardening bias $\beta=0.5$ and extremity of items $\alpha = 0.3$. 
    Panels \textbf{A} and \textbf{B} show example trajectories of attitudes (solid line) under limited and abundant information. Additionally, the panels include the full vector of beliefs as held by the agent at each point in time, colored by whether the agent currently believes (green) or does not believe (gray) the item. The argument consulted at any specific timepoint is enlarged, indicating whether the item was believed (dark green circle) or rejected (grey cross). 
    Panel \textbf{C} shows evolution of the probability distribution of attitudes of the agent when information is limited, and Panel \textbf{D} shows the results when information is abundant.
    }
    \label{fig:persuasive-arguments}
\end{figure*}

To check the generality of the previous results, we represent information abundance and hardening confirmation bias in a second, fundamentally different model.
In Persuasive Argument Theory~\cite{myers1982polarizing,Maes2013a,Shamon2019changing,Banisch2021argument}, agents exchange arguments in favor or against a topic. When receiving an argument, an agent decides to believe it or not according to its current attitude about the topic. Believing a new positive argument increases the agent's attitude while believing a negative argument decreases it. The version of Persuasive Argument (PA) model proposed by Banisch and Shamon~\cite{Banisch2023biased} includes a hardening confirmation bias that is in line with experimental observations. When the agent's attitude about the topic becomes stronger on one side, the agent accepts arguments on this side and rejects arguments from the other side with a higher probability. Our second model builds on this work.   

\subsection*{Single agent model}

Like in the BC agent model, we assume a single agent that, at any point in time, chooses at random to consult one among $N$ available items of information. These items of information are called arguments in the previous PA models. Now, $N$ is assumed even and for $i \in \{1,...,N/2\}$ item $i$ is in favor of the topic ($c_i = 1$, called a positive item) and for $i \in \{N/2+1,...,N\}$ item $i$ is against the topic ($c_i = -1$, negative item). Moreover, while with the BC model the extremity of items is drawn uniformly, in the PA model, we assume that the extremity $0 < \alpha \leq 1$ is the same for all items. Therefore the attitude of item $i$ is $c_i \alpha$. 
The agent holds a belief $b_i \in \{0, 1\}$ about each item $i \in \{1,.., N\}$. This belief vector indicates which items the agent takes into account when forming its opinion. Initially, we assume that the agent is neutral and does not hold any beliefs (i.e., all $b_i = 0$). When picking item $i$, the agent decides to believe it ($b_i = 1$) or not ($b_i = 0$).

\subsection*{Hardening confirmation bias}

Following~\cite{Banisch2023biased} who introduced confirmation bias in PA models, the probability to believe an item depends on the current balance of beliefs $\overline{b}$ defined as the sum of beliefs $b_i$ weighted by their corresponding attitude ($c_i \alpha$):
\begin{equation}\label{eq:balanceofbeliefs}
    \overline{b} =  \sum_{j =1}^{N} b_j c_j \alpha.
\end{equation}

The probability to believe any item $i$ is defined as
\begin{equation}\label{eq:probBelief}
    P(b_i = 1) = \frac{1}{1 + \exp(-\beta  \overline{b} c_i)}
\end{equation}

When the number of believed items of one sign increases, the probability to believe items of this sign increases (and the probability to believe items of opposite sign decreases). This expresses a hardening confirmation bias, which strength is ruled by parameter $\beta$. Using data from a survey experiment,~\cite{Banisch2023biased} estimated different $\beta$'s for different topics, and found an average value of $\beta \approx 0.5$. In our computational experiments, we keep this value.

\subsection*{Limited vs abundant information}

Like in the BC model, we are interested in comparing the evolution of the agent's attitude when information is limited and when it is abundant. In our tests, we consider again the case $N=10$ items (5 positive and 5 negative) and the case of $N = 100$ items (50 positive, 50 negative).

\subsection*{Results}

Fig.~\ref{fig:persuasive-arguments} presents example trajectories and probability distributions of agent attitudes (see details in Materials and Methods section for how to derive the attitude from the belief vector). Panels A and B show the fluctuations of attitudes in limited and abundant information cases, and the full belief vectors containing positive and negative items over time. In Panel A, the attitude derived from limited information keeps fluctuating around a neutral value, while beliefs about the positive and negative items regularly change. The agent in Panel B quickly believes more and more positive items and reaches an almost maximally extreme and stable attitude.

Panels C and D of Fig.~\ref{fig:persuasive-arguments} show the distributions of attitudes over time for the same parameters, for the equivalent of an infinity of trajectories in a Markov model (see Materials and Methods). The distribution is non-zero on isolated points because the balance of beliefs takes values in the finite set: $\{-N\alpha/2,...,+N\alpha/2\}$. The distribution is thus represented by bars expressing the probability of the attitude corresponding to an exact value of the balance of beliefs.


The difference of evolution of the distribution of attitudes between limited and abundant information is even more striking than with the BC model. When information is limited, attitudes remain moderate, while they extremize strongly when information is abundant. 

This result is easy to understand. Indeed, in this model, the stronger the attitude on one side, the more likely it is to become even stronger if new arguments can strengthen it. Therefore, if the strongest attitude, corresponding to a balances of beliefs equal to $\pm N \alpha / 2$, is extreme, with a strong confirmation bias, reaching and keeping this attitude over time is very likely. On the contrary, if the strongest attitude is moderate, with a weak confirmation bias, then it is unstable as believing arguments on the other side is likely. Hence, the likelihood to keep the strongest attitude is low.  

Overall, for a given value of the argument extremity $\alpha$, if the number $N$ is small enough for the strongest attitude to remain moderate, then the attitudes remain moderate. On the contrary, if $N$ is large enough for the strongest attitude to be extreme, the evolution of attitudes tends to the extremes after many item consultations.
In particular, the attitude can become extreme even when the items are very moderate ($\alpha$ is small) if the number of items $N$ is large enough (see examples in the SI). This is a noticeable difference with the BC model in which extremization requires consulting extreme items. 

\section*{Discussion}

This study reveals that information abundance---as brought about by the proliferation of accessible content on the internet---may play a primary role in attitude extremization. We show that even the most balanced information feed may drive users into dogmatist and extreme attitudes, when consulting a lot of information. 
Indeed, both our models yield the same qualitative result: when the agents consult abundant information, hardening confirmation bias is very likely to drive them to firm and extreme positions. 
Moreover, the process of extremization is similar in both models. 
When information is abundant and diverse, hardening confirmation bias creates an asymmetry: it is easier to become more extreme than to become more moderate. With this asymmetry, gradually, a random drift towards extreme attitudes takes place. This process is less likely to occur when information is limited, because there are then less opportunities to become gradually more extreme. 

Our work comes with a series of limitations.
Firstly, the models make strong simplifying assumptions in order to show clear results. The existence of the processes that the models reveal remains speculative while not confirmed experimentally~\cite{Deffuant2023c}.  
Secondly, even if their existence is empirically confirmed, these processes can be in competition with other dynamics of extremization. For instance, research  suggests that the American political system favors extreme attitudes~\cite{Enns2021,Broockman2019} and echo chambers could combine their extremization effects with the ones of information abundance. 



Keeping these limitations in mind, we conclude this discussion by considering polarization and online media regulation, in the light of our results.

The general American public is not polarized on political issues~\cite{Lelkes2016}, while the elites are increasingly polarized~\cite{McCarty2019}. 
Our study sheds new light on these observations. 
Indeed, for any given topic, only a minority of agents in a large population regularly and actively seeks information about the topic (activists and people who are heavily involved in the topic). If this minority is small, the extremization of these agents' attitudes may be difficult to observe in population level surveys.   
The SI also proposes a theoretical example illustrating that the extremization of a $10\%$ minority is not distinguishable from standard uncertainty in the distribution of attitudes of the whole population. 

However, regulating online media in order to mitigate the extremization due to information abundance looks particularly challenging.
Indeed, in our models, extremization takes place even with rigorously balanced information. As this extremization process is rooted in a fundamental cognitive bias, the only means for its regulation is to limit information abundance, which is very controversial and faces very serious practical obstacles. Interestingly, as previously stressed, the BC model differs from the PA model in this respect. The BC model indeed predicts that limiting extreme content would limit attitude extremization, while the PA model predicts that this would only delay it (see SI for details). This further motivates experimental research assessing these models.  
Overall, the consequences of our work are two-sided. On the one hand, the revealed extremization process only applies to people consulting abundant information about the issue at stake. As these people are generally a minority, this process is unlikely to cause mass polarization by itself. On the other hand, deep root of confirmation bias in our cognitive functioning makes this process particularly challenging to mitigate, especially if---as predicted by the PA model---extremization takes place even when consulting only moderate items of information. 


\section*{Materials and Methods}

\subsection*{Agent with Bounded Confidence}

Each item $i \in \{1,...,N\}$ holds attitude $c_i$ uniformly drawn in $[-1, 1]$. At each iteration $t$, the agent consult a randomly chosen agent $i$. The agent's attitude $a_t$ is updated to $a_{t+1}$ with the following equation:

\begin{equation}
a_{t+1} = 
    \begin{cases}
        \frac{1}{2}(a_t + c_i) & \mbox{ if  } |a_t - c_i| \leq \epsilon(a_t)\\
        a_t & \mbox{ otherwise}.
    \end{cases}
\end{equation}
where $\epsilon(a_t)$ denotes the confidence bound associated with a specific attitude. The confidence bound is a hard threshold for ignoring the item. As agents become more extreme in their attitudes, this confidence bound hardens (as observed experimentally). The function specifying the decreasing confidence bound is:
\begin{equation}
\epsilon(a) = \epsilon_M \exp(- \beta a^2)    
\end{equation}
where parameter $\beta$ rules this decrease, hence the hardening of confirmation bias. 

\subsection*{Agent consulting Persuasive Arguments}

At each iteration, the agent consults a randomly chosen item $i \in \{1,...,N\}$, of attitude of sign $c_i = \pm 1$ and of extremity $\alpha$. The probability to believe an item of sign $c_i$ is:
\begin{align}
    P(b_i = 1) = \frac{1}{1 + \exp(-\beta \overline{b} c_i)},
\end{align}
where the balance of beliefs $\overline{b}$ denotes the sum of an agent's belief vector as defined in Eq.~\ref{eq:balanceofbeliefs}. We then map this balance of beliefs onto an attitude $a \in [-1, 1]$. We define the attitude as the probability to believe a positive item minus the probability to believe a negative item. Formally, the attitude $a_t$ is defined as:
\begin{equation}\label{eq:normOpinion}
    a_t = \frac{1 - \exp(-\beta \overline{b}_t)}{1 + \exp(-\beta \overline{b}_t)}
\end{equation}

We have tested and confirmed the robustness of our results to an other definition of attitude which is independent from parameter $\beta$---the normalized balance of beliefs---described in the SI.


\subsection*{Markov model of agent consulting Persuasive Arguments}

To precisely compute the average evolution of attitude of the PA agent, we use a Markov model. This model defines the evolving probability $\rho_t(b_p, b_n)$  that the agent believes $b_n$ negative and $b_p$  positive items. Initially, the couple $b_p = 0$ and $b_n = 0$ has a probability 1 (and all the other couples have a probability 0). Then, the probabilities evolve at each iteration, by taking into account the probability of transition from a couple to another.  These transition probabilities between couples are a function of the balance of beliefs $\overline{b} =(b_p - b_n)$ and parameter $\beta$, ruling confirmation bias hardening. By iteratively updating the probability distribution $\rho_t(b_p, b_n)$ across all possible couples, the Markov model yields in one run the exact average result over an infinity of simulation runs of the agent model. The SI provides details on the transition probabilities and on the algorithm of the Markov model.
 



\section*{References}
\begin{multicols}{2}

\bibliography{references}

\section*{Acknowledgements}
Guillaume Deffuant acknowledges funding from European Union’s Horizon-Widera program under the TED4LAT project (Number 101079206). Marijn Keijzer acknowledges IAST funding from the French National Research Agency (ANR) under the Investments for the Future (Investissements d'Avenir) program, grant ANR-17-EURE-0010. Sven Banisch acknowledges funding from the European Union’s Horizon Europe program under grant agreement No 101094752 (SoMe4Dem -- Social Media for Democracy.
\end{multicols}

\end{document}